# Spectral and Dual-Energy X-ray Imaging for Medical Applications

Erik Fredenberg [a,*]

[a] *Philips Research, Knarrarnäsgatan 7, 164 85 Kista, Sweden*

**Abstract**

Spectral imaging is an umbrella term for energy-resolved x-ray imaging in medicine. The technique makes use of the energy dependence of x-ray attenuation to either increase the contrast-to-noise ratio, or to provide quantitative image data and reduce image artefacts by so-called material decomposition. Spectral imaging is not new, but has gained interest in recent years because of rapidly increasing availability of spectral and dual-energy CT and the dawn of energy-resolved photon-counting detectors. This review examines the current technological status of spectral and dual-energy imaging and a number of practical applications of the technology.

*Keywords*: X-ray imaging; Spectral imaging; Dual energy; Computed tomography; Mammography; Radiography

## 1. Introduction

J. C. Maxwell developed the theory of electromagnetic waves during the 1860's and predicted an infinite range of frequencies [1]. It was soon assumed that visible, infrared, and ultraviolet light was part of the electromagnetic spectrum, and the discovery of radio waves by H. Hertz followed during the late 1880's [2]. W. C. Röntgen discovered x radiation in 1895 and noted its similarity to light [3], but because of the particle properties of the new rays, their connection to the electromagnetic spectrum remained unclear. Quantum interpretations of the photo-electric effect [4] and of Compton scattering [5, 6], the dominating interaction effects for medical x-ray imaging in general and crucial for energy-resolved imaging, were important pieces in the puzzle that led up to L. de Broglie's explanation of the wave-particle duality in 1924 [7], which helped to kick off quantum mechanics.

---

* Corresponding author. Tel.: +46-702-766130; e-mail: erik.fredenberg@philips.com.

The controversy around the nature of x radiation during the early 1900's did not hinder practical use of the rays. W. H. and W. L. Bragg pioneered spectroscopy around 1913 [8], and, in parallel, there was a rapid development of medical applications of x rays [9], evidenced, for instance, by the first use of contrast media in 1904 [10]. Subsequently, the medical community started to show interest in spectral effects in x-ray imaging, and, for instance, the effect of monochromatic x rays for radiography was discussed at the 32$^{nd}$ Annual Meeting of the Radiological Society of North America (RSNA) in 1946 [11]. In 1953, Jacobson presented a method inspired by x-ray absorption spectroscopy to measure the concentration of iodine in x-ray images using exposures at two different energies [12].

The advent of computed tomography (CT) during the 70's spurred a renewed interest in energy-resolved imaging. The potential for measuring atomic numbers with exposures at two different voltage levels was proposed already by G.N. Hounsfield in his landmark CT paper [13], and the technology evolved rapidly for contrast-enhanced imaging [14, 15] and unenhanced imaging [16–20]. The concept of unenhanced imaging was subsequently also extended to projection imaging [17].

Following the early conceptual development of energy-resolved imaging, limitations in available technology for long held back wide-spread clinical use. Specifically, temporal differences between multiple exposures [21], including patient, cardiac, and respiratory motion, as well as variation in contrast-agent concentration, posed problems, in particular for CT. In recent years, however, two fields of technological breakthrough have spurred a renewed interest in energy-resolved imaging. Firstly, incremental development of source and detector technologies for several decades finally reached a tipping point that enabled clinical introduction of single-scan energy-resolved CT systems by several major manufacturers [22–24]. Starting in 2006, this disruptive increase in availability and commercial interest has generated a large number of clinical applications for energy-resolved imaging. Secondly, and in parallel, detector technology is going through a paradigm shift with the advent of photon-counting detectors in clinical practice. The first commercial photon-counting system was introduced for mammography in 2003 [25], and CT systems are at the verge of being feasible for routine clinical use [26]. Photon-counting detectors have the potential to overcome many of the limitations of previous techniques for energy-resolved imaging, for instance in the number of energy levels, and open up for a vast number of additional clinical applications.

Jacobson used the term "dichromography" for the technology to acquire images at two different x-ray energy spectra [12]. Other notations with a "hard coded" number of energy levels followed, and "dual energy" is still the dominating terminology for energy-resolved x-ray imaging. The terms "spectral imaging" and "spectral CT" started to appear about ten years ago to acknowledge the fact that photon-counting detectors have the potential for measurements at a larger number of energy levels [27–31]. In this review, "spectral imaging" and "energy-resolved imaging" will be used as general umbrella terms for different kinds of energy-resolved x-ray imaging, but the more specific "dual-energy imaging" will be used in parallel.

In the following, the development and current status of clinical spectral and dual-energy x-ray imaging from a technology and applications perspective will be outlined. Photon-counting technology may be of particular interest because of its promising future prospects and is also the topic of another review in this volume, concerning the photon-counting Medipix chip [32]. Energy-resolved imaging in synchrotron facilities, though interesting, is not within the scope.

## 2. Background of spectral and dual-energy imaging

### 2.1. Spectral image acquisition

An energy-resolved imaging system probes the object at two or more photon energy levels. In a generic imaging system, the projected signal in a detector element at energy level $\Omega \in \{E_1, E_2, E_3 \dots\}$ is

$$n_\Omega = q_\Omega \int \Phi_\Omega(E) \exp[-\mu_1(E)t_1 - \mu_2(E)t_2 - \mu_3(E)t_3 \dots] \, \Gamma_\Omega(E) \mathrm{d}E, \qquad (1)$$

where $q$ is the number of incident photons, $\Phi$ is the normalized incident energy spectrum, and $\Gamma$ is the detector response function (adapted from Ref. [33], Eq. 3.15). Linear attenuation coefficients and integrated thicknesses for a number of materials that make up the object are denoted $\mu$ and $t$, which attenuate the x-ray beam according to Lambert-Beers law. Two conceivable ways of acquiring spectral information are to either vary $q \times \Phi$ with $\Omega$, or to have $\Omega$-specific $\Gamma$, referred to in the following as, respectively, incidence-based (c.f. Sec. 3.1) and detection-based methods (c.f. Sec. 3.2 and Sec. 3.3).

Most elements appearing naturally in human bodies have low atomic number and therefore lack absorption edges in the diagnostic x-ray energy range. The two dominating x-ray interaction effects in this range, Compton scattering and the photo-electric effect, can then be assumed to be smoothly varying functions with separable and independent material and energy dependencies. The linear attenuation coefficients can hence be expanded as [16]

$$\mu = a_{PE} \times f_{PE}(E) + a_C \times f_C(E). \tag{2}$$

The photo-electric material coefficient ($a_{PE}$) depends strongly on atomic number ($\propto Z^4$), whereas the Compton coefficient ($a_C$) has a weaker atomic number dependence ($\propto Z$), but depends on the electron density, which is in general proportional to mass density. The photo-electric effect depends strongly on photon energy ($f_{PE} \propto E^{-3}$), whereas the Compton-scattering energy dependence ($f_C$) is relatively flat over the diagnostic energy range and is often modelled by the Klein-Nishina cross section [6]. Figure 1 shows the linear attenuation of typical breast tissue consisting of 20% glandular tissue and 80% adipose tissue, separated into photo-electric and Compton components. Note that Eq. (1) and Eq. (2) treat scattering as absorption, i.e. efficient scatter rejection is assumed.

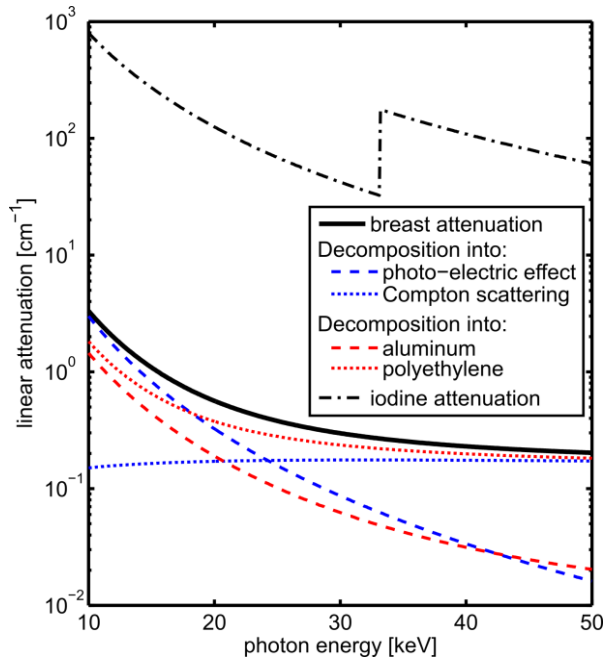

Figure 1: Linear attenuation as a function of photon energy. The attenuation of a 20%-density breast is decomposed into photo-electric + Compton bases (blue) and aluminum + polyethylene bases (red). The linear attenuation of iodine illustrates the effect of a contrast material with a K edge. Attenuation data are from the XCOM data base [34], and breast composition is from Ref. [35].

In contrast-enhanced imaging, one or several high-atomic-number contrast agents with K absorption edges in the imaged energy range may be present in the body. Figure 1 shows an example of iodine with a K edge at 33.2 keV. K-edge energies are material specific, which means that the energy dependence of the photo-electric effect is no longer separable from the material properties, and an additional term can be added to Eq. (2) in order to describe the linear attenuation [36]:

$$\mu = a_{PE} \times f_{PE}(E) + a_C \times f_C(E) + \sum a_K \times f_K(E), \qquad (3)$$

where $a_K$ and $f_K$ are the material coefficients and energy dependencies of contrast-agent material $K$, and the sum is over the range of administered contrast agents.

### 2.2. Energy weighting

A simple sum over the energy bins ($n = \sum n_\Omega$) yields a conventional non-energy-resolved image, but because x-ray contrast varies with energy, a correspondingly weighted sum ($n = \sum w_\Omega \times n_\Omega$) is optimal from a contrast-to-noise-ratio aspect [37]. The benefit of so-called energy weighting (c.f. Sec. 4.1) is most prominent in energy regions where the photo-electric effect is non-negligible (low energies and / or high-$Z$ materials) because of its strong energy dependence, and optimal weights follow approximately the energy dependence of the photo-electric effect ($E^{-3}$) [38]. For the example of breast-tissue in Figure 1, the intersection between the photo-electric and Compton cross sections occurs at 24 keV, above which the efficiency of energy weighting is limited.

### 2.3. Material decomposition

Besides energy weighting, a second approach to make use of the energy information is to treat the acquisitions according to Eq. (1) as a system of equations with material thicknesses as unknowns, a technique broadly referred to as material decomposition (c.f. Sec. 4.3 to Sec. 4.9 for clinical applications). System properties ($q_0, \Phi, \Gamma$) and linear attenuation coefficients ($\mu_1, \mu_2, \mu_3, ...$) are assumed to be known, either explicitly (e.g. by modelling) or implicitly (by calibration).

In CT, implementing material decomposition post reconstruction (image-based decomposition) [39] does not require coinciding projection data and is therefore feasible for all methods of acquiring energy-resolved data (c.f. Sec. 3), but the decomposed images may suffer from beam-hardening artefacts. A polychromatic beam traversing an object is hardened because low-energy photons are preferentially absorbed, and the attenuation as a function of position is therefore not identical from different projection angles, but rather a function of traversed thickness, object composition, and photon energy spectrum. These effects are difficult to take into account in 3D reconstruction and lead to artefacts, which are passed on to the material-decomposed images because the reconstruction algorithm is generally non-reversible (c.f. Sec. 4.2). Applying material decomposition directly in projection space instead (projection-based decomposition) [16], similar to projection imaging, can in principle eliminate beam-hardening artefacts because the decomposed projections are quantitative and do not vary with acquisition parameters.

For projection imaging and projection-based decomposition in CT (and in the absence of K-edge contrast agents), the limited number of independent energy dependencies according to Eq. (2) means that the system of equations can only be solved for two unknowns, and measurements at two energies ($|\Omega| = 2$) are necessary and sufficient for a unique solution of $t_1$ and $t_2$ [17]. Materials 1 and 2 are referred to as basis materials and are assumed to make up the object; any other material present in the object will be represented by a linear combination of the two basis materials. Figure 1 shows the decomposition of breast tissue into aluminium-polyethylene bases. The basis-material representation can be readily converted to images showing the amounts of photo-electric and Compton interactions by invoking Eq. (2), also illustrated in Figure 1, and further to images of effective-atomic-number and electron-density distributions by utilizing the known material dependencies of the photo-electric and Compton cross sections [16]. Moreover, as the basis-material representation is sufficient to fully describe the linear attenuation energy dependence of the object, it is possible to calculate monochromatic images, i.e. images showing the object as it would appear at any photon energy (c.f. Sec. 4.2).

An important exception to the limit of two unknowns is if, on top of the energy-resolved data, additional information is available for the decomposition. In particular, if the total tissue thickness can be assumed to be known, another equation is effectively added to the system, i.e. $t_1 + t_2 = t_{tot}$, where $t_{tot}$ is known. For instance, within the compressed breast region in mammography, iodine [40], as well as tumour tissue or cyst fluid [41], can be separated from adipose and glandular tissue by assuming slow thickness variations. Other studies have used external thickness measurement devices [42]. Equivalently, in reconstructed CT data the voxel size is known; realizations of Eq. (1) at a full angular range adds orthogonal data to the system of equations, which enables image-based three-material decomposition [43]. Tomosynthesis, on the other hand, does not provide a full range of projection images and the added spatial data is only partially orthogonal and depending on spatial frequency in the reconstructed image [44].

In contrast-enhanced imaging, additional unknowns may be added to the system of equations according to Eq. (3) if one or several K absorption edges are present in the imaged energy range [36]. With one K-edge contrast agent, measurements at three energies ($|\Omega| = 3$) are necessary and sufficient for a unique solution, two contrast agents can be differentiated with four energy bins ($|\Omega| = 4$), et cetera. This technique is often referred to as K-edge imaging (c.f. Sec. 4.6) [27]. A traditional method of contrast-enhanced material decomposition is to linearize the system of equations made up by realizations of Eq. (1) by taking the log (separable energy dependencies of the system properties and the linear attenuation coefficients are assumed) and solving by Gaussian elimination, a technique commonly referred to as dual-energy subtraction [14, 45]. More advanced methods are common today, for instance based on maximum-likelihood algorithms for over-determined systems [28].

It is interesting to note that spectral imaging can be treated as a task-dependent optimal combination of the energy bins [40]. Energy weighting falls out as a special case when the task function is constant at all spatial frequencies, i.e., corresponding to a delta function. Material decomposition is optimal for more advanced imaging tasks, for instance, to detect iodine-enhanced lesions over a cluttered background.

## 2.4. System optimization

The performance of spectral and dual-energy imaging applications increases with increased signal difference, i.e. reduced overlap, between the energy bins [46, 47]. For energy weighting as well as material decomposition without contrast enhancement, large separation in mean energy between the energy bins is generally beneficial because, away from any absorption edges, linear attenuation decreases monotonically with energy. For incidence-based methods, large spectral separation is accomplished by large differences in kV between the exposures and / or heavy filtering of the high-energy beam [48, 49], restricted towards the low-energy end by patient dose, and towards the high-energy end by the available photon flux. For detection-based methods, the spectral separation is limited by the energy resolution of the detector and the choice of tube voltage and filtering [45]; a large span of the spectrum increases spectral separation but also dose towards lower energies.

For K-edge imaging the situation is somewhat opposite because, in the presence of absorption edges, the attenuation does not decrease monotonically with energy [14]. With a singular rise in attenuation at the K absorption edge, the signal increases with the measured energies narrowing on either side of the edge and decreases with spectral separation. Photon-counting detectors, in which the energy bins can be set arbitrarily, are most suited for K-edge imaging [50], even though K-edge filters [14], or even x-ray optics [30], can be used in incidence-based methods to shape the spectrum around the edge.

Most studies on optimization of spectral and dual-energy imaging systems consider spectral separation, dose, and photon-flux, but also more advanced schemes, including, for instance, motion artefacts have been proposed [46].

## 3. Technologies and methods

### 3.1. Incidence-based methods

The most straight-forward way of obtaining spectral information is to run several standard image acquisitions at different tube voltage settings, possibly in combination with different filtering. This was the technique used in the very early studies on energy-resolved projection imaging and CT, and is still widely used in projection imaging [51], and CT [52, 53]. As filtering and tube voltage selection is not intrinsically restricted, this technique does not in general (as has sometimes been assumed) increase radiation dose compared to detection-based methods [54]. The drawbacks are rather related to temporal differences between the exposures [21], including patient, cardiac, and respiratory motion, as well as variation in contrast-agent concentration, which, in particular for CT, for long limited the usefulness of the technique. With the subsequent development of spiral CT and multi-row detectors, these limitations are now less pronounced, and the technique is still available in commercial CT systems.

The introduction of dual-source CT by Siemens in 2006 was mainly meant to improve the temporal resolution for non-spectral cardiac CT [55], but also enabled dual-energy imaging by virtually eliminating motion artefacts between two exposures [56], for which spectral cardiac CT is the most prominent example [57]. The Siemens dual-source system has two source-detector pairs placed at approximately 90° angle relative to each other, which acquire images at two different exposure settings, typically 80 and 140 kVp. Voltage, current, and filtration can, however, be set independently for the two x-ray tubes, which has facilitated optimization in subsequent versions by for instance increasing tube voltage for obese patients and adding additional tin filtration to the high-energy beam for improved spectral separation [49, 58]. The field-of-view of the high-energy detector is smaller than standard [55], which poses difficulties for obese patients, but this is not an intrinsic limitation of the technique. Scatter between the two source-detector pairs is, on the other hand, a challenge intrinsically associated with this particular setup [59]. Also, as images are acquired along different ray paths for the two source-detector pairs, projection-based material decomposition is challenging, and the images need to be reconstructed prior to decomposition.

Figure 2 a) shows a schematic, a photo, and approximate spectral distributions for the Siemens dual-source system as well as for examples of commercial systems representing the other techniques that will be described in this section. The spectral distributions for the CT systems were simulated according to Ref. [60], but with attenuation data from Ref. [34] and incident spectra from Ref. [61]. The energy resolution of the photon-counting mammography system was adopted from Ref. [31].

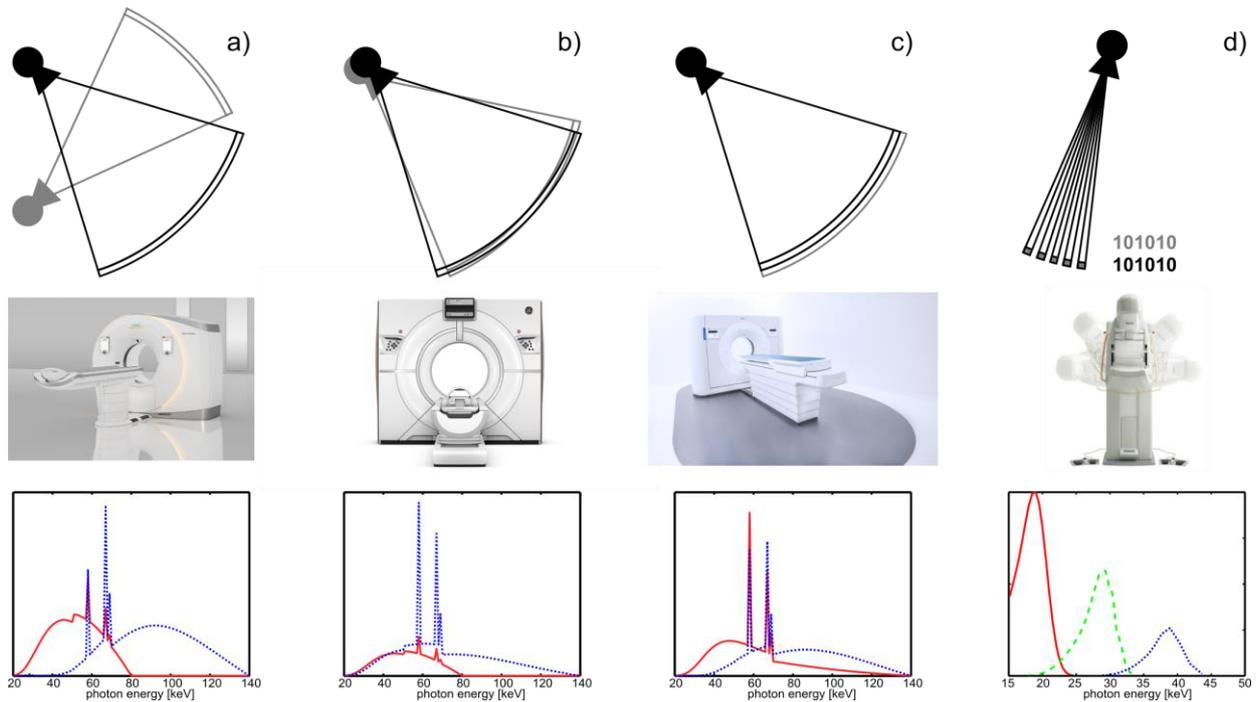

Figure 2: Technologies for spectral and dual-energy imaging in commercial systems, exemplified by a) the Siemens Somatom Drive dual-source CT (80 kV / 140 kV + 0.4 mm additional Sn filtration, photo from www.siemens.com/presse), b) the GE Revolution CT with rapid kV switching (80 kV / 140 kV), c) the Philips IQon CT with sandwich detector (garnet scintillator + gadolinium oxysulphide scintillator), d) the Philips MicroDose photon-counting mammography system. The CT spectra were simulated according to Ref. [60]. For the mammography system, the energy resolution shows the response to monochromatic impulses at 20, 30, and 40 keV, fitted to a detector model [31]. The multi-exposure technique is not exemplified in this figure because it can be accomplished by virtually any x-ray system.

An alternative to the dual-source implementation of incidence-based dual-energy CT is to rapidly switch between exposure settings [62, 63]. This technology was introduced commercially in 2008 by General Electric (GE) and is denoted fast kV switching [23, 64]. In the GE system, the x-ray tube voltage is switched, typically between 80 and 140 kV, several thousand times per second in order to acquire images at high- and low-energy spectra through virtually identical ray paths (c.f. Figure 2 b)). Compared to dual-source CT, current challenges are mainly related to the high switching rate; it is difficult to introduce different filtering for the two exposures, the difference in tube voltage between the exposures is limited, the tube current cannot be changed between the exposures, even though the dose per projection can be altered by the relative exposure times.

*3.2. Integrating spectral detectors*

Very early clinical work on energy-resolved imaging focused on the incidence-based techniques because detectors with energy resolution were mainly developed for spectroscopy with heavy filtering (for instance based on crystal refraction) and therefore extremely dose inefficient. Nevertheless, the breakthrough came relatively soon with the introduction of so-called sandwich detectors [45, 65–67]. Sandwich detectors consist of two (or more) detector layers, where the top layer preferentially detects low-energy photons and the bottom layer detects a filtered, and therefore harder, spectrum. The filtering comprises in some implementations just the top detector layer, whereas in some cases a thin metal filter, most often a copper filter, is introduced between the detector layers, which increases spectral separation between the bins, but reduces dose efficiency.

A sandwich detector for chest imaging has been commercially available from Fuji. The Fuji system is a computed radiography (CR) system with two phosphor plates and a 0.8-mm sandwiched copper filter [68, 69]. A CT system based on a sandwich detector was introduced commercially by Philips in 2015 (Figure 2 c)) [24, 70]. The detector consists of an yttrium-based garnet scintillator layer on top of a gadolinium oxysulphide layer. A thin front-illuminated photodiode is mounted on the side of the scintillator, and hidden beneath the anti-scatter grid in order not to reduce the active area of the detector. There is no added filtration between the scintillator layers, which maximizes dose efficiency, but spectral separation may be lower than for some incidence-based methods [60]. There is no change in scan protocol for spectral imaging, and the spectral information is therefore available from every scan, which has practical / workflow advantages [71]. The two energy levels measured by the detector represent identical ray paths, which facilitates projection-based material decomposition.

*3.3. Photon-counting spectral detectors*

A presently ongoing paradigm shift for detection-based methods and for spectral imaging in general is the advent of photon-counting detectors for medical x-ray imaging [72]. As opposed to conventional detectors, which integrate all photon interactions over a certain time interval, photon-counting detectors are fast enough to register single photon events. Similar to other spectral detectors, photon-counting detectors detect all energy levels simultaneously, which means that there are no temporal changes between the bins that may introduce artefacts, there is no special protocol needed so spectral information is in principle available from all acquisitions, and projection-based material decomposition is facilitated in CT. However, unlike any other methods described above, the number of energy bins and the spectral separation are not determined by physical properties of the system (number and properties of detector layers, sources etc.) but are determined by the detector electronics, which increases the degrees of freedom and the potential efficiency. Even though many photon-counting systems today suffer from technical limitations, the intrinsic limitations are relatively few, and photon-counting detectors have a huge potential for the future. Also, the fact that the spectral separation occurs at a very late stage in the imaging chain makes it possible to compensate for inefficiencies in the detection process by sorting out electronic noise or detector scatter that would otherwise deteriorate the spectral performance. It can be noted that, except for improved spectral performance, photon-counting detectors offer additional advantages compared to many conventional detectors, such as improved spatial resolution [73].

Photon-counting detectors have been used in nuclear medicine for decades, but their introduction to transmission imaging was relatively late, mainly as a result of the higher flux that leads to pileup. It was therefore not a coincidence that the first commercial photon-counting application is a mammography system [74], in which high expectations on spatial resolution result in small detector elements and therefore relatively low count rates. The so-called MicroDose mammography system was introduced by Sectra Mamea in 2003 (acquired by Philips in 2011) [25], and spectral imaging was launched on this platform in 2013 [40]. The MicroDose system is based on an array of silicon strip detectors in a multi-slit configuration that is scanned across the object to acquire an image (Figure 2 d)) [31]. The multi-slit geometry provides high intrinsic scatter rejection [75], which is particularly advantageous for spectral imaging as Compton scattering can be treated as an absorption mechanism without further assumptions. The technology is currently making its way into breast tomosynthesis [76].

Photon-counting image receptors are roughly made up of a sensor and electronics. The sensor material is usually a solid-state semiconductor, such as silicon [31, 74, 77–80]. The advantages of silicon include high charge-collection efficiency, ready availability of high-quality high-purity silicon crystals, and established methods for test and assembly driven by the semiconductor industry [81]. The main challenge is the relatively low photo-electric cross section, which limits the detection efficiency and leads to a large fraction of Compton scatter in the detector. The low detection efficiency is often addressed by arranging the silicon wafers edge on [82, 83]. Compton scatter degrades the energy response because the full photon energy is not be deposited in a single detector channel and compensation may be necessary, for instance, by a separate energy threshold to sort out Compton scattered events [81]. A key characteristic of a photon-counting spectral detector is its energy resolution. The energy resolution of the MicroDose detector ranges from $2.0 - 2.3$ keV standard deviation in the range 20-40 keV, and is shown in Figure 2 d) for monochromatic impulses at 20 keV, 30 keV, and 40 keV, derived from a fitted detector model [31]. Newer silicon detectors have improved the energy resolution to less than 2 keV in the 40-120 keV interval [78].

Several research groups and commercial companies are investigating cadmium telluride (CdTe) and cadmium-zinc telluride (CZT) as sensor materials for photon-counting CT [27–29, 84–87], and for projection imaging [88, 89]. The higher atomic number of these materials result in higher absorption and less detector scattering, but, on the other hand, the higher K-fluorescent yield leads to degraded spectral response and cross-talk between detector elements [90, 91]. Also, manufacturing of macro-sized crystals of these materials poses practical challenges, and the crystals generally suffer from lattice defects and impurities that lead to charge trapping. Charge trapping limits the charge-collection efficiency for individual events [90, 92], and may also cause long-term polarization effects (build-up of space charge) [93], which reduces energy resolution and detector speed.

Other solid-state materials, such as gallium arsenide [94], and mercuric iodide [95], are currently quite far from clinical implementation, but may be expanding in the future. Gas detectors have also been investigated [96], but gas is more difficult to handle than solid materials and have limited absorption efficiency.

A bias voltage is applied over the sensor, so that when a photon interacts in the material, charge is released and drifts as electron-hole pairs towards the anode and cathode respectively, where it is collected by electrodes. The electronics of current photon-counting detectors are almost exclusively implemented in application-specific integrated circuits (ASICs), with each electrode being connected to parallel channels in the ASIC [31, 86, 89, 97–99]. There may be slight differences in the exact ASIC implementation, but in general [92], each channel comprises an amplifier and a shaper, which convert the charge to a pulse with a height proportional to the energy of the impinging photon. The pulse height is measured by comparators, referred to as energy thresholds, which are followed by corresponding counters. The counters yield the sum of all events, which generate a pulse and exceeds the corresponding threshold value, and the detected photons are divided into bins according to energy by differentiating the counter values. Up to eight energy bins are currently being implemented [78]. A low-energy threshold below the expected incident spectrum is used to prevent electronic noise from being counted, but electronic noise will still be added onto the pulse height and to some extent influence the energy resolution [100]. Low-energy thresholding may also be used to sort out Compton scattered events that would otherwise degrade energy resolution [81]. Figure 3 shows the basic principle of a photon-counting spectral detector according to the description above.

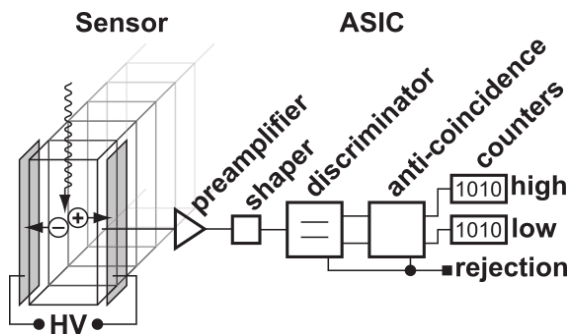

Figure 3: Basic principle of a photon-counting spectral detector. Charge is released when a photon interacts in the sensor and drifts under the influence of a bias voltage to the electrodes. A preamplifier and a shaper convert the charge to a pulse with a height that is proportional to the energy of the impinging photon, and the pulses are divided into bins according to energy by comparators. Anti-coincidence may be implemented in the ASIC to discriminate against charge sharing, but in most practical such functionality is not available.

The main challenge of photon-counting detectors is pulse pileup [92, 101–103], which occurs when more than one photon interacts in a single detector channel within the time window for pulse detection, the so-called dead time. Pileup results in lost counts and reduced energy resolution because several pulses are counted as one, with a height that is a superposition of the true pulse heights, heavily depending on the distance between the events. Pileup will always be present in photon-counting detectors because of the Poisson distribution of x-ray photons, but the speed of detector electronics has increased substantially during the past ten years, and CT count rates are coming within reach [104].

Another challenge of photon-counting detectors is cross-talk between adjacent channels, induced, for instance, by fluorescent x-rays or so-called charge sharing, which occurs when charge from a single photon interaction is collected by neighbouring electrodes and therefore detected as several events with distributed photon energy [82, 90]. Such cross talk can be mitigated by anti-coincidence logic in the ASIC [92], which sorts out pulses in adjacent detector channels within a certain time window. The size of this time window is a compromise between efficiency of the technique and the probability of discriminating against true quasi-coincident counts, so called chance coincidence. Chance coincidence together with increased complexity of the electronics reduces the maximum count-rate and increases power consumption, which are the main drawbacks of this technique. Coincident counts can be excluded all together, added to a high-energy bin (because charge sharing occurs with higher probability for high-energy photons) [31], or treated with more advanced schemes that include summation of the pulse height from adjacent detector channels and a probability-based localization of the impulse, such as implemented in the Medipix chip [105].

One way of reducing the count rates per channel and simultaneously providing additional energy resolution for photon-counting detectors is to segment the sensor with several electrodes in the depth direction [81, 106], such that electrodes at larger depth detect harder spectra, thus in a way combining photon-counting and sandwich detector technology. Figure 4 shows a segmented silicon-strip detector for photon-counting spectral CT [78]. Another way of improving energy resolution is to discard photons detected in every second bin so that the spectral separation between the bins that are used for spectral analysis increases [107]. This approach, somewhat similar to introducing filters in sandwich detectors, increases quantum noise, but can potentially improve spectral performance.

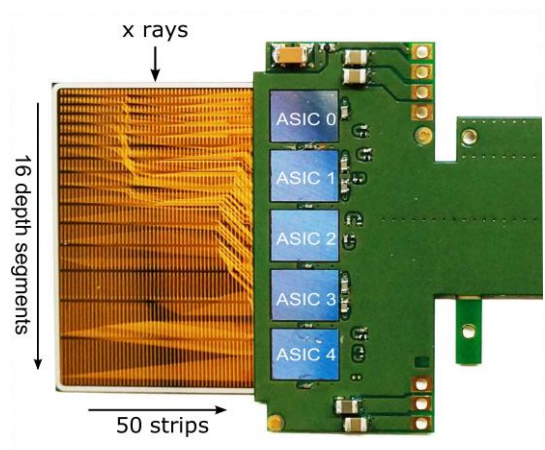

Figure 4: Silicon-strip detector for photon-counting spectral CT with the silicon wafer oriented edge-on. 50 strips are segmented into 16 depth segments and read out by 5 ASICs with 160 channels each. Image courtesy Prof Mats Danielsson and Dr.Cheng Xu.

## 4. Applications

### 4.1. Energy weighting

Energy weighting was pioneered by Tapiovaara and Wagner [37], and has subsequently been refined for projection imaging, including mammography [38, 88, 108–111]. In CT, energy weighting can either be applied on the projections [112], or on the reconstructed images [113]. Energy weighting improves the pixel-to-pixel contrast-to-noise ratio (CNR) by assigning a higher weight to low-energy photons. Thus, the dose efficiency is effectively improved, which enables a higher CNR at a constant patient dose, or a lower dose at a constant CNR. The effect follows as a logical consequence of the fact that the x-ray contrast drops with increasing photon energy, which holds true in energy regions away from absorption edges and where the photoelectric effect is non-negligible, including most x-ray imaging applications without contrast agents.

Several studies have investigated energy weighting at the limit of infinite energy resolution, which results in a CNR improvement of approximately 9% in mammography [38, 110], 6% for light elements (wax) in CT [114], and 24-80% for heavier elements in CT [113, 114], compared to photon counting without energy resolution (the improvement compared to non-photon-counting systems is larger). However, studies with realistic energy resolution report CNR improvements in the range 1-5% [108, 109, 115] for mammography, 4% for light elements (wax) in CT [114], and 7-30% for heavier elements in CT [113, 114]. In particular, Berglund et al modified a clinical spectral photon-counting mammography system and was able to improve the CNR of clinical images by 2.2-5.2%, which translates to a potential dose reduction at a constant CNR in the range of 4.5-11% [111]. Even though these CNR improvements are relatively moderate, they come for free if imaging protocols are not altered for spectral acquisitions, such as in incidence-based methods (sandwich and photon-counting detectors).

### 4.2. Virtual monochromatic images

As noted in Sec. 2.3, the full object linear attenuation as a function of energy can be determined with measurements at just a few energy levels because of the limited number of x-ray interaction effects, and energy-resolved imaging therefore allows for reconstruction of virtual monochromatic images. Such monochromatic display is useful for optimizing the CNR to a certain imaging task, analogous to energy weighting for soft tissue as described above. For instance, an important imaging task in neuroradiology is to distinguish between grey and white brain matter in CT images because minute variations in contrast give an early indication of infarction. It has been found that the CNR between grey and white matter increases in virtual monochromatic images at low energies because of increased soft-tissue contrast compared to higher energies, and the optimal virtual energy for this task has been determined to be in the range 65-75 keV [116, 117].

Further, the CNR in contrast-enhanced imaging can be optimized by choosing an energy of the virtual monochromatic image slightly above an absorption edge of the contrast agent [118]. One application of this strategy is found in CT angiography (CTA), where monochromatic images at 50 keV (for iodine with a K edge at 33.2 keV) enable reduction of the iodine load to minimize toxicity or the risk of allergic reactions when characterizing pulmonary embolism [119]. In oncology, monochromatic images can improve tumour conspicuity. Pancreatic cancer, for instance, is one of the leading causes of cancer mortality in the Western world and is notoriously difficult to visualize in conventional contrast-enhanced CT. Subjective image-quality and CNR are, however, improved in virtual monochromatic images at $40 - 55$ keV for iodinated contrast agents [120, 121], with reported CNR improvements of up to 7 times [120].

Another application of virtual monochromatic images is to reduce beam-hardening artefacts in CT. When a polychromatic beam traverses the object it is hardened because low-energy photons are preferentially absorbed, and the attenuation as a function of position is therefore not identical from different projection angles, but rather a function of traversed thickness, object composition, and photon energy spectrum, which poses problems for the reconstruction algorithm.

Another application of virtual monochromatic images is to reduce artefacts caused by highly-attenuating objects in CT, which interfere with, for instance, imaging tasks in the vicinity of artificial metal objects, such as in the positioning of pedicle screws [122], or for evaluation of interfaces between prostheses, bone, and surrounding tissue [123, 124]. Artefacts also impede many imaging tasks close to the scull (in the posterior fossa and in the subcalvarial space) [116, 125]. Virtual monochromatic images are typically generated from material decomposed images and beam-hardening artefacts are therefore mitigated as discussed in Sec. 2.3. Nevertheless, highly-attenuating objects also generate artefacts from photon starvation, which cannot be eliminated but are somewhat compensated for at higher virtual energies because of higher penetration. Optimal virtual energies for these types of imaging tasks are therefore typically found in the region around or above 100 keV [116, 122–125], as a compromise between reduced artefacts and soft-tissue or iodine contrast.

It is clear that projection-based decomposition is substantially more efficient in reducing beam-hardening artefacts than is image-based decomposition because the reconstruction algorithm is generally non-reversible (c.f. Sec. 2.3) [126, 127]. In addition, incidence-based spectral imaging may yield higher CNR improvements in soft-tissue differentiation than rapid kVp switching because the latter requires interpolation between adjacent acquisitions to allow for projection-based decomposition [116].

### 4.3. Virtual non-contrast image

Contrast agents have been applied since the dawn of x-ray imaging when H. Rieder noted that imaging of the gastrointestinal tract was facilitated by ingestion of bismuth. (The so-called "Rieder meal" was, however, replaced by barium contrast agents within a few years because of the toxicity of bismuth salts [128].) Except for gastrointestinal radiology, contrast agents have been developed for a range of applications, including examination of lung ventilation (for instance by xenon inhalation) and examination of the vascular system (for instance by iodinated contrast agents). High-atomic-number contrast agents have substantially higher x-ray absorption, and therefore higher contrast, than soft tissue because of the strong $Z$ dependence of the photo-electric effect.

In contrast-enhanced procedures involving intravenous administration of a contrast agent, a non-contrast image is often required prior to contrast injection in order to separate morphology and contrast agent, either by visual comparison or by a weighted subtraction of the two images, so-called temporal subtraction. Energy-resolved imaging enables decomposition of a single contrast-enhanced acquisition into contrast and non-contrast images (under the assumptions listed in Sec. 2.3).

The non-contrast image of a spectral or dual-energy procedure, often referred to as a virtual unenhanced or virtual non-contrast (VNC) image, can save dose to the patient in cases where the virtual non-contrast image is sufficiently good that the additional non-contrast acquisition can be omitted. For instance, characterization of renal masses with CT requires a non-contrast image to assess the distribution of contrast agent relative the anatomy. Using a VNC image as the baseline may reduce radiation dose by 19-35% [129, 130].

Another application of VNC images is to remove contrast-agent residuals in unenhanced images following contrast-enhanced procedures. For instance, detection of intracerebral haemorrhage secondary to thrombolysis for treatment of ischemic stroke is facilitated by removing remaining iodine contrast [131], which is of particular value when stenting and administration of anti-coagulants become necessary.

In radiotherapy, exact delineation of the lesion is crucial and contrast enhancement is often necessary, but the contrast agent, which is not present at the time of treatment, may render erroneous calculations of the dose distribution. Using a VNC image for dose calculations yields distributions similar to those calculated in true unenhanced images [132].

### 4.4. Improved quantification of contrast agent

The contrast image of a spectral or dual-energy procedure displays the distribution of the contrast agent, possibly overlaid (e.g. in colour) on the virtual non-contrast image for mapping to anatomy, and can be made quantitative to measure actual concentrations. In principle, the same result can be obtained with temporal subtraction, but this technique suffers heavily from artefacts caused by patient motion between contrast injection and until the contrast agent has spread through the vascular system. The value of the contrast image was realized early; it was in fact the main motivation behind the first discussions about energy-resolved projection imaging [12] and CT [13].

Tumours generally exhibit increased permeability and retention of intravenous contrast agents, caused by angiogenesis associated with the irregular growth [133]. A range of studies have investigated energy-resolved mammography [40, 134–136], and breast tomosynthesis [137, 138], for minimizing adipose-glandular contrast (anatomical clutter) and thereby improving visibility of iodine-enhanced lesions. Multi-phase studies have demonstrated the ability to measure contrast-agent kinetics, which provides further functional information [139]. In a recent study [51], the performance of contrast-enhanced dual-energy mammography was not significantly different from that of breast magnetic resonance imaging (MRI), which indicates that the technology is coming close to offer a fast and low-cost alternative to MRI for diagnostic workup and treatment monitoring. In dual-energy CT, the contrast image has been used for cancer diagnostics, including lymph node staging and therapy response monitoring in lung cancer patients [140].

In cardiovascular imaging, CT perfusion imaging (CTP) is emerging for diagnosing coronary artery stenosis and myocardial ischemia with the benefit of retrieving functional information about the contrast-agent distribution in addition to visualizing morphology. Energy-resolved imaging improves CTP by reducing beam-hardening artefacts caused by bone tissue and vast amounts of contrast agent in the large vessels around the heart, but also by separating morphological and functional information (e.g. in coloured overlays of iodine distribution), thus improving visualization [57, 127, 141, 142].

In thoracic imaging, CT angiography (CTA) is the reference standard for diagnosing acute pulmonary embolism. Energy-resolved imaging adds functional perfusion information by improved visualization of the contrast-agent distribution [143], which potentially increases detection, particularly for small peripheral emboli [144], even though the benefit is still somewhat unclear. Further, combining iodine enhancement for perfusion and xenon enhancement for ventilation imaging may add diagnostic value [145].

*4.5. Differentiation between contrast agent and other attenuating structures*

X-ray absorption by naturally occurring calcium often obscures iodine contrast, and energy-resolved imaging may improve these contrast-enhanced procedures by differentiating between iodine and calcium. For instance, bone contrast can be supressed for better visualization of the contrast-enhanced vascular system in CT angiography. Dual-energy based bone removal has been rated superior to non-energy-resolved techniques in studies of the head and neck region [146], and in whole-body CT [147].

A more specific application is to improve characterization of atherosclerotic plaque to assess the risk of rupture, which is the major cause of myocardial infarction, by differentiating between iodinated contrast agent and calcifications, as well as between the lipid core of the plaque and other soft tissue. Studies of this application using dual-source CT have only been able to show limited improvement compared to conventional CT [148, 149], but photon-counting CT [150], possibly in combination with novel contrast agents [151], may overcome the limitations of present clinical systems.

Cystic lesions in the kidneys is a common finding in abdominal CT imaging. Simple benign cysts generally do not accumulate contrast agent, but can nevertheless exhibit increased contrast due to high protein content or haemorrhage, which may obscure contrast-enhancement of malignant soft tissue. A phantom study has demonstrated that material decomposition with spectral CT enables simple cysts to be unambiguously distinguished from contrast-enhanced lesions, which may facilitate risk assessment for hyperattenuating cystic lesions [152].

*4.6. Novel contrast agents and K-edge imaging*

Most studies of contrast-enhanced spectral and dual-energy imaging have used iodine, which is a well-established contrast agent for x-ray imaging, but the K edge of iodine at 33.2 keV is not optimal for all applications. In mammography, the iodine K-edge energy might be too high, resulting in low statistics for measurements above the edge, and / or tube voltages above optimal values. In CT, on the other hand, the edge might be too low and fall below the detected spectrum. Some studies on CT imaging have used gadolinium, which is an MR contrast agent with a K edge at 50.2 keV [153]. Except for having a more well-positioned K-edge energy, gadolinium can be used for patients who are hypersensitive to iodine. Recent advances in nanotechnology have made it possible to produce contrast agents based on nanoparticles from a range of materials. For instance, one study investigated silver with a K edge at 25.5 keV as a contrast agent for mammography [154]. Other studies have investigated zirconium (K edge at 18.0 keV) for mammography [155], and gold (K edge at 80.7 keV) for CT [156]. Except for the possibility of further optimizing the K edge energies with a larger range of materials, some contrast agents can be targeted [151], which opens up possibilities for molecular imaging.

K-edge imaging enables decomposition of more than two materials by employing the additional interaction effect provided by K-shell absorption in heavy elements used for contrast agents [27–29, 81]. In the presence of K absorption edges in the imaged energy range, material decomposition at two energy levels, such as in most clinical systems today, is not quantitative because the dimensionality of the attenuation energy dependence is higher than two (c.f. Sec. 2.3). Photon-counting detectors, however, enable a larger number of energy levels, which can be arbitrarily placed above and below any absorption edge in order to quantitatively probe the contrast agent [27], and, for instance, separate contrast agent, calcified plaque and stent material in cardiovascular imaging [29].

Further, photon-counting detectors with more than three energy bins open up for multi-agent imaging, i.e. to quantitatively measure the distribution of more than one contrast agent in the body, for instance iodine and gadolinium [28]. A recent phantom study demonstrated that it may be feasible to improve visibility of polyps in photon-counting colonography using orally administered iodine and intravenously administered gadolinium as contrast agents [157]. Another study used targeted gold nanoparticles to characterize macrophage burden in atherosclerotic plaque, while simultaneously using non-specific iodinated contrast agent to image vascularity [151]. Multi-agent imaging can also be applied to reduce the number of acquisitions in multi-phase imaging; by administering two or more contrast agents at different time points, the distribution of these respective contrast agents represent different uptake phases at a certain acquisition time point [73].

### 4.7. Characterization of lesions and abnormalities without contrast enhancement

In addition to contrast-enhanced imaging, material decomposition can be applied for quantitative measurements on body tissues. For instance, in screening mammography, injection of contrast agent is not desirable. The holy grail of unenhanced spectral and dual-energy mammography has been to improve the detection of solid tumours over background tissue, but this has so-far been difficult because of the minute difference in attenuation between glandular tissue and tumour tissue [109, 158]. Improving detection of micro calcifications, an early sign of malignancy in breast imaging, has also been investigated [159], but pixel noise may instead be a limiting factor for detecting the small calcifications. In chest imaging, on the other hand, a similar but more successful application is to supress the contrast of overlapping ribs for easier detection of pulmonary nodules [68, 160–162].

Spectral imaging has been applied in mammography to differentiate between cysts and solid tumours [76, 163], which has the potential to reduce the amount of recalls from screening and thereby limit stress and anxiety for the patient and cost for society. A pilot study reported a specificity of approximately 50% at a sensitivity level of 1% for discriminating between solids and cysts [41]. Related to cyst-tumour differentiation is differentiation between benign and malignant micro calcifications in the breast, which is initially promising, at least for larger clusters of calcifications [164].

Differentiation between different types of renal calculi (stones) is an application of energy-resolved CT that helps select appropriate treatment; the less common uric acid stones are preferentially treated with urinary alkalinization whereas the more common non-uric acid stones (containing e.g. calcium oxalate, calcium phosphate, struvite, or cystine) often require removal by external shock wave lithotripsy or invasive procedures. Non-uric acid stones contain high-atomic-number elements that are not present in uric acid stones and the characterization with energy-resolved CT is therefore relatively straight forward, in particular for larger stones, and exhibit close-to 100% accuracy [53, 165–167]. Differentiation between types of non-uric acid stones has additional diagnostic value and is possible, but with a lower accuracy of around 80% [166, 167]. The application was first proposed for dual-exposure techniques early in the history of CT [168], and has been refined for state-of-the art systems [165–167], even though dual exposure with image processing to account for patient motion is still a feasible option for relatively stationary organs, such as the kidneys [53].

Related to the classification of renal calculi is detection of uric acid crystal deposition in joints to differentiate gouty arthritis from other conditions with similar symptoms. Early diagnosis of gout is crucial for effective treatment and to avoid associated renal and cardiac disease, but the current reference standard, joint aspiration, has limited sensitivity and is sometimes technically challenging to perform. Dual-energy CT can detect uric acid crystals in joints with sensitivity and specificity in the 80% - 90% range [169, 170], and can also be used for treatment monitoring by measuring the urate volume with low inter- and intra-reader variability [170].

The reference standard for diagnosing so-called bone bruises (bone marrow edema) is MRI, partly because calcium obscures the visualization with conventional CT. Dual-energy CT can remove calcium contrast from overlying trabecular bone by material decomposition (separating bone mineral, yellow bone marrow, and red bone marrow), which enables detection of posttraumatic bone bruises with sensitivity and specificity of approximately 86% and 95%, respectively [171].

*4.8. Characterization of normal tissue*

Energy-resolved imaging can be used to measure the breast composition in terms of adipose and glandular tissue [172–176], referred to as breast density, which is related to the risk of developing breast cancer, the efficiency of screening mammography, and can be used for treatment monitoring. Spectral breast-density measurement has proven more consistent than non-spectral methods [177].

Measurement of bone-mineral density, an independent risk factor for fractures and all-cause mortality, was one of the first investigated applications of energy-resolved CT [19, 63] and has also been explored with modern approaches [178–180]. Compared to dual-energy x-ray absorptiometry, energy-resolved CT may improve the accuracy because of the possibility to exclude areas with artefacts and interfering anatomy as well as the cortical bone, which is less metabolically active and therefore less sensitive to changes in bone mineral density than the inner trabecular bone. In case the spectral data comes for free, such as in detection-based methods, opportunistic osteoporosis screening can be carried out in all patients undergoing CT examinations with a mean error in the calcium hydroxyapatite concentration of less than 6% over a range of standard protocols [180]; bone-mineral analysis alone might not justify a CT scan.

Proton therapy is emerging as an alternative to conventional radiotherapy with the benefit of a highly focused dose distribution towards the end of the finite proton range (the so-called Bragg peak), which reduces the dose to healthy tissue. Accurate calculation of proton stopping power is, however, crucial in order to make use of the improved precision offered by proton therapy. Effective-atomic-number and electron-density image pairs generated with energy-resolved CT have the potential to increase the precision of this calculation substantially [181–183].

### 4.9. Foreign objects

High-contrast foreign objects in the body, originating for instance from invasive procedures, can be difficult to distinguish from contrast enhancement. Hence, besides reduction of metal artefacts as discussed in Sec. 4.2, spectral imaging may improve imaging tasks involving foreign objects by material decomposition, such as in one case where a suspected bleeding could be excluded and determined to be a suture [184].

So-called virtual autopsies conducted with postmortal dual-energy CT examinations may facilitate detection and characterization of bullets, knife tips, and glass or shell fragments, which are of forensic importance to determine the cause of injury [185]. Further, characterization of implants of various types may aid in identifying unknown corpses.

### 4.10.  Summary of clinical applications

Table 1 lists the clinical applications that have been discussed above, sorted into the categories energy weighting, virtual monochromatic images, and material decomposition. Further, the cited studies are, as far as possible, sorted on medical specialty and body region.

The division into categories in Table 1 and the sections above is somewhat artificial and the borders between applications is not always so sharp; a single clinical investigation may involve more than one of these applications.

Further, despite the breadth of specialties and body regions, it should be noted that the citations in this review are still merely examples and by no means cover the full range of published studies.

**Table 1:** Clinical applications of spectral and dual-energy imaging with example references, stratified on medical specialty and body region.

| Technology / Application | | Benefit | Example studies: Medical specialty / Body region |
|---|---|---|---|
| Energy weighting | | Improved CNR | General: [37, 88, 112–114]<br>Oncology / Breast: [38, 88, 108–111, 115] |
| Virtual monochromatic images | | Improved soft-tissue and contrast-agent CNR. Reduced beam hardening and photon starvation artefacts. | General: [118, 126]<br>Cardiology / Cardiovascular: [127]<br>Pulmonology / Thoracic: [119]<br>Neurology / Head, neck: [116, 117, 122]<br>Oncology / Abdomen, head, neck: [120, 121, 125]<br>Orthopaedics / Musculoskeletal: [123, 124] |
| Material decomposition | Virtual non-contrast image | Reduced radiation dose. Improved workflow. Removal of iodine staining. Radiotherapy dose calculations. | Neurology / Head, neck: [131]<br>Urology, oncology / Abdomen: [129, 130]<br>Oncology / Radiotherapy: [132] |
| | Contrast-enhancement of lesions | Improved detection and characterization. | Oncology / Breast, chest: [40, 51, 134–140] |
| | Perfusion, angiography | Reduced beam hardening. Improved visualization. | Cardiology / Cardiovascular: [57, 127, 141, 142]<br>Pulmonology / Thoracic: [143–145] |
| | Contrast-agent differentiation | Plaque characterization. Bone removal in angiography. | General: [147]<br>Neurology / Head, neck: [146]<br>Cardiology / Cardiovascular: [148–151]<br>Urology / Abdomen: [152] |
| | Optimized contrast agents | Improved visualization of contrast agent | General: [154–156] |
| | Targeted contrast agents | Molecular imaging | Cardiology / Cardiovascular: [151] |
| | K-edge imaging | Improved differentiation. Multi-agent imaging. | General: [28, 81, 73]<br>Cardiology / Cardiovascular: [27, 29, 151]<br>Gastroenterology / Abdomen: [157] |
| | Characterization of unenhanced abnormalities | Improved detection of lesions. Differentiation between benign and malignant lesions. Differentiation between renal stones. Diagnosis of gout and bone marrow edema. | Oncology / Breast, chest: [41, 68, 76, 109, 158–164]<br>Urology / Abdomen: [53, 165–168]<br>Orthopaedics / Musculoskeletal: [169–171] |
| | Normal tissue characterization | Measurement of breast density and bone mineral density. Calculation of proton stopping power. | Oncology / Breast, radiotherapy: [172–177, 181–183]<br>Orthopaedics / Musculoskeletal: [19, 63, 178–180] |
| | Foreign objects | Improved detection and characterization of foreign objects. | Cardiology / Cardiovascular: [184]<br>Pathology: [185] |

## 5. Conclusions and Outlook

There is a rapid increase in the number of clinical applications of spectral and dual-energy imaging, including two- and three-dimensional imaging, spanning the entire body (breast, chest, head and neck, cardiovascular, abdominal, musculoskeletal, etc.), and ranging over a multitude of specialties (oncology, neurology, urology, orthopaedics, pathology, etc.). This development is aided by the spread and increasing availability of commercial systems for energy-resolved CT and projection imaging, including photon-counting mammography. Photon-counting CT is close to being feasible for routine clinical use. So far, the majority of studies on energy-resolved CT are based on the dual-source technique because of early market introduction, but a larger variety can be expected in future studies, utilizing the benefits of the respective techniques.

X-ray imaging has a strong position in morphological imaging with generally high spatial resolution and short exposure times (high spatiotemporal resolution). Energy-resolved imaging is able to strengthen this position by reducing artefacts and improving CNR. In addition, material decomposition provides information that is complementary to the conventional image and opens up for functional imaging and physiological characterization, an area traditionally dominated by other modalities such as MRI and SPECT.

Further, energy-resolved imaging can be made quantitative, which is in line with the strong trend towards personalized medicine and precision medicine [186, 187]. These two related fields target tailoring of care flows to individual patients or patient groups in order to maximize value for patients and society. Quantitative information from material decomposition facilitates implementation of precision medicine in the diagnosis stage, for instance by personalized screening protocols, and in the therapy stage, for instance by treatment monitoring and planning. In parallel, surgery is also moving towards higher precision with minimally invasive techniques, in which quantitative image guidance is crucial. One can speculate that we will see more applications of energy-resolved imaging in this field, which, for instance, take advantage of quantitative imaging and the ability to better visualize foreign objects in the body by reduced beam hardening and differentiation from contrast agents.

The added dimension in energy-resolved imaging causes an increase in the amount of generated data, which may be a future challenge as healthcare systems are already troubled by vast amounts of imaging data, not only from the technical perspective (storage, handling etc.), but also from the staffing perspective (someone has to do the reading). It is reasonable to believe that methods for data analytics, artificial intelligence, and machine learning, which grow within several fields in medicine [188], will be matured for energy-resolved imaging applications and assist in comprehending and taking full advantage of the data.

## Acknowledgements

Special thanks are extended to Dr Hans Bornefalk, Dr Klaus Erhard, Dr Matthijs Kruis, and Dr Ewald Roessl, all with Philips, as well as Prof Mats Danielsson and Dr Cheng Xu, Royal Institute of Technology, Stockholm, Sweden, and Dr Håkan Almqvist, Karolinska University Hospital, Solna, Sweden, for valuable discussions, insightful feedback, proof reading, and various important contributions to this review. The author would also like to thank the Editor of this issue and the editor appointed Referee for a constructive peer-review process, which has clearly improved the quality of the paper.